\documentclass[11pt]{article}
\usepackage{moriond,epsfig}

\def\MyPhoto{\psfig{figure=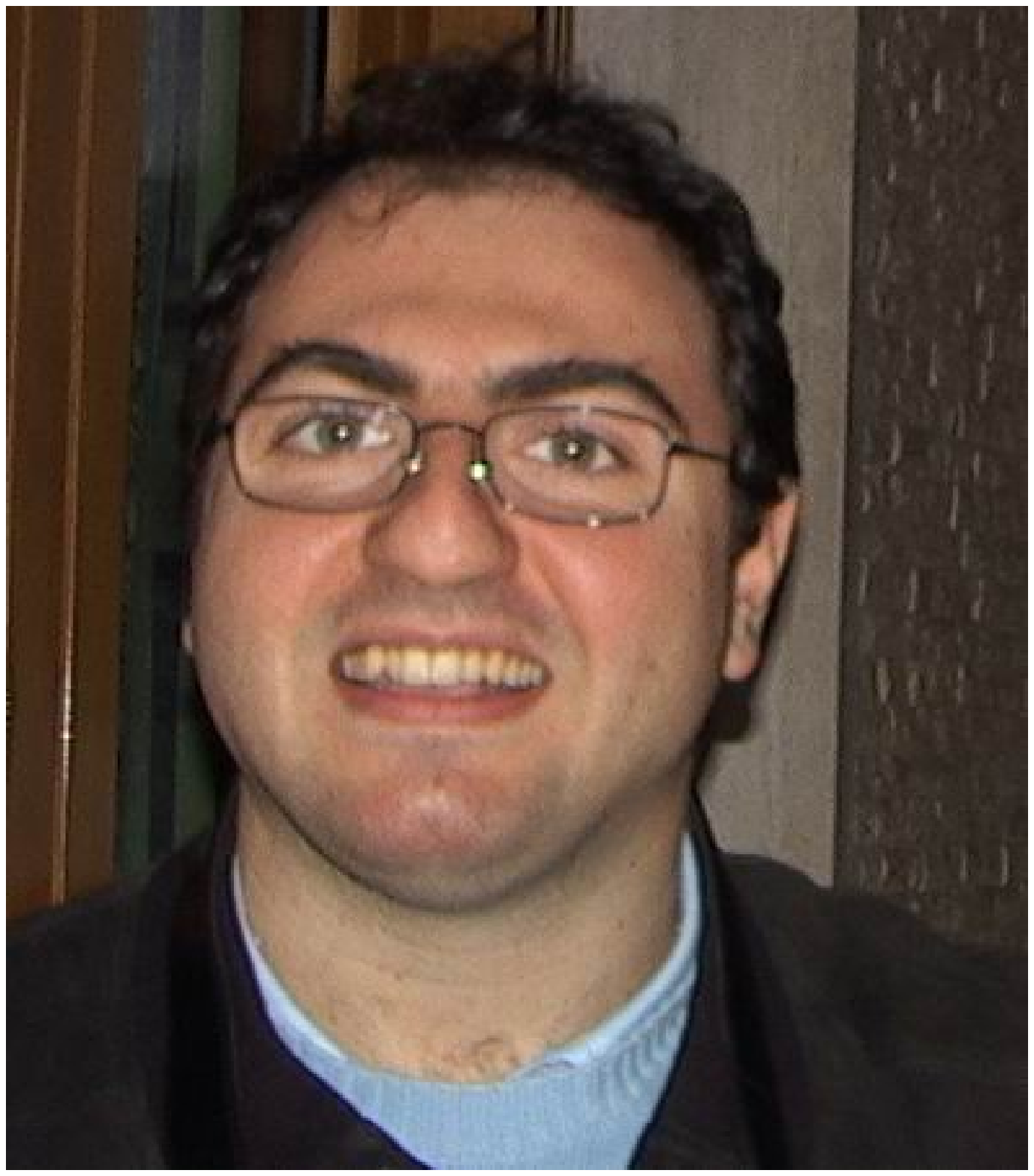,width=35mm}}

\def\Journal#1#2#3#4{{#1} {\bf #2}, #3 (#4)}

\def\PLB{{\em Phys. Lett.}  B}
\def\PRL{\em Phys. Rev. Lett.}

\def\PRC{{\em Phys. Rev.} C}

\def\be{\begin{equation}}
\def\ee{\end{equation}}
\def\bea{\begin{eqnarray}}
\def\eea{\end{eqnarray}}

\begin{document}
\vspace*{4cm}
\title{HIGGS PHYSICS AND BEYOND THE STANDARD MODEL AT CMS/ATLAS}

\author{ N. DE FILIPPIS }

\address{Laboratoire Leprince Ringuet, $\mathrm{\acute{E}}$cole Polytechnique - IN2P3/CNRS \\ 
91128, PALAISEAU, France}

\maketitle\abstracts{Prospective searches about Higgs physics and beyond the Standard Model are
presented for the CMS and ATLAS experiments. Possible excesses of events in real data could be an indication
of the existence of new particles, even with few hundred $\mathrm{pb^{-1}}$ of
integrated luminosity. In this paper the focus is on the current analyses strategies and on
the potential both for a discovery and/or for an exclusion of the Standard Model Higgs boson in
the main decay channels. The searches for some supersymmetric and exotic particles predicted
by several theoretical models are also discussed.}

\section{Introduction}

The field of high energy physics is approaching an important period
of its history with the start of the operations of the Large Hadron
Collider (LHC) at CERN, the world's largest and highest-energy particle
accelerator. The LHC will collide opposing beams of protons or lead
ions, each carrying energies per nucleon up to 2.76 TeV. LHC started to
operate with the injection of first beams in the beam pipe in fall
2008. The LHC has been built with the purpose of exploring new frontiers of
particle physics, giving evidence of the existence of the Higgs
boson and/or a wide spectrum of new particles predicted by
supersymmetry and exotic models.

In general, the experiments at the LHC could provide answers or
ingredients to answer some of the most {fundamental open questions}
in particle physics, such as: the reality of the Higgs mechanism for
generating gauge bosons and fermions masses, the problem of the hierarchy between
the electroweak gauge boson scale and the Grand Unification or Planck scale,
the existence of a supersymmetry which implies that the
known Standard Model (SM) particles have supersymmetric partners, the existence of
extra dimensions as predicted by various models inspired e.g. by
string theory.

CMS and ATLAS are the two general purpose experiments built at the LHC aimed to
provide answers to those fundamental questions.
Prospective studies have been performed over the last years in the
physics groups of the CMS and ATLAS collaborations to optimize
strategies for the search of the Higgs boson(s), of the
supersymmetric particles and of some exotic particles predicted by
several models, at the center-of-mass energies of the LHC collider,
both with low and high integrated luminosity.

\section{Prospective searches at CMS/ATLAS}

The search for Higgs and supersymmetric particles has been the major guide to
define the detector requirements and performance that are detailed in Ref.~\cite{CMSdet} and Ref.~\cite{ATLASdet} for
CMS and ATLAS.

Detailed simulations of the detector closest to the real
experimental set-up with miscalibration/misalignment conditions at
start-up luminosity have been used in the CMS and ATLAS studies.

Advanced Monte Carlo physics generators has been used for signal and
background simulation with the estimation of NLO QCD and electroweak
corrections.

\subsection{Searches for Standard Model Higgs}\label{subsec:higgssm}

Direct searches for the SM Higgs particle at the LEP
$\mathrm{e^+e^-}$ collider have led to a lower mass bound of
${m_{\rm H} > 114.4 \, \mathrm{GeV/c^2}}$ at 95\% C.L.~\cite{higgsLEP}. On-going direct searches at the Tevatron
$\mathrm{p\bar{p}}$ collider by the D0 and CDF experiments set
constraints on the production cross-section for a SM-like Higgs boson in a mass range extending up to about $200 \,
\mathrm{GeV/c^2}$ and allow to exclude his existence~\cite{higgsTevatron} with mass
between 160 and 170 $\mathrm{GeV/c^2}$.

The main production mechanisms for SM Higgs particle at
LHC are the gluon–-gluon fusion
mechanism, the associated production with W/Z bosons, the weak
vector boson fusion processes and the associated Higgs production
with heavy top or bottom quarks, as detailed in Ref.~\cite{theory}. The gluon fusion mechanism
dominates especially at low Higgs mass and the cross section at NLO
is in between 0.1 and 50 pb depending on the Higgs mass; the cross
section of Higgs production via the vector boson fusion is generally
one order of magnitude lower with respect to gluon fusion while the
other contributions are much less important.

SM Higgs couples to fermions, gauge bosons and to
itself. In the low mass region (namely $\mathrm{m_{H}<130 \, GeV/c^2}$) the dominant decay is in $\mathrm{b
\bar{b} }$ with a branching ratio between 60 and 90 $\%$; $\mathrm{H
\rightarrow \tau^+ \tau^-, \, c \bar{c}, \, \gamma \gamma}$
contribution to the total width is less that few \%. In the high
mass range the decay channels $\mathrm{H\rightarrow WW^{(*)}}$ and $\mathrm{H\rightarrow
ZZ^{(*)}}$ play the main role given a clear signature of multi
leptons in the final state.

A prospective analysis about the $\mathrm{H \rightarrow WW \rightarrow ll
\nu \nu}$ decay chain was performed both in CMS~\cite{PAS_CMS1} and
in ATLAS~\cite{ATLAS1}. The signature consists of two isolated high
momentum leptons and missing energy related to the neutrinos
escaping the detection. No hard jet in the central region of the
acceptance is expected and it is not possible to reconstruct the
Higgs mass peak due to the neutrinos. The main background
comes from $\mathrm{t \bar{t}}$ and di-boson events, di-leptons a la Drell-Yan, $\mathrm{tW}$
and W+jets events in the topologies including two leptons in the
final state.

The analysis mainly consists of selecting events with high
transverse momentum leptons and sufficient missing energy; a central jet veto strategy is
used to select events with no hard jet in central rapidity region
and the angular correlation between the leptons coming from Higgs
decays is used as a discriminating observable.

\begin{figure}[t]
\psfig{figure=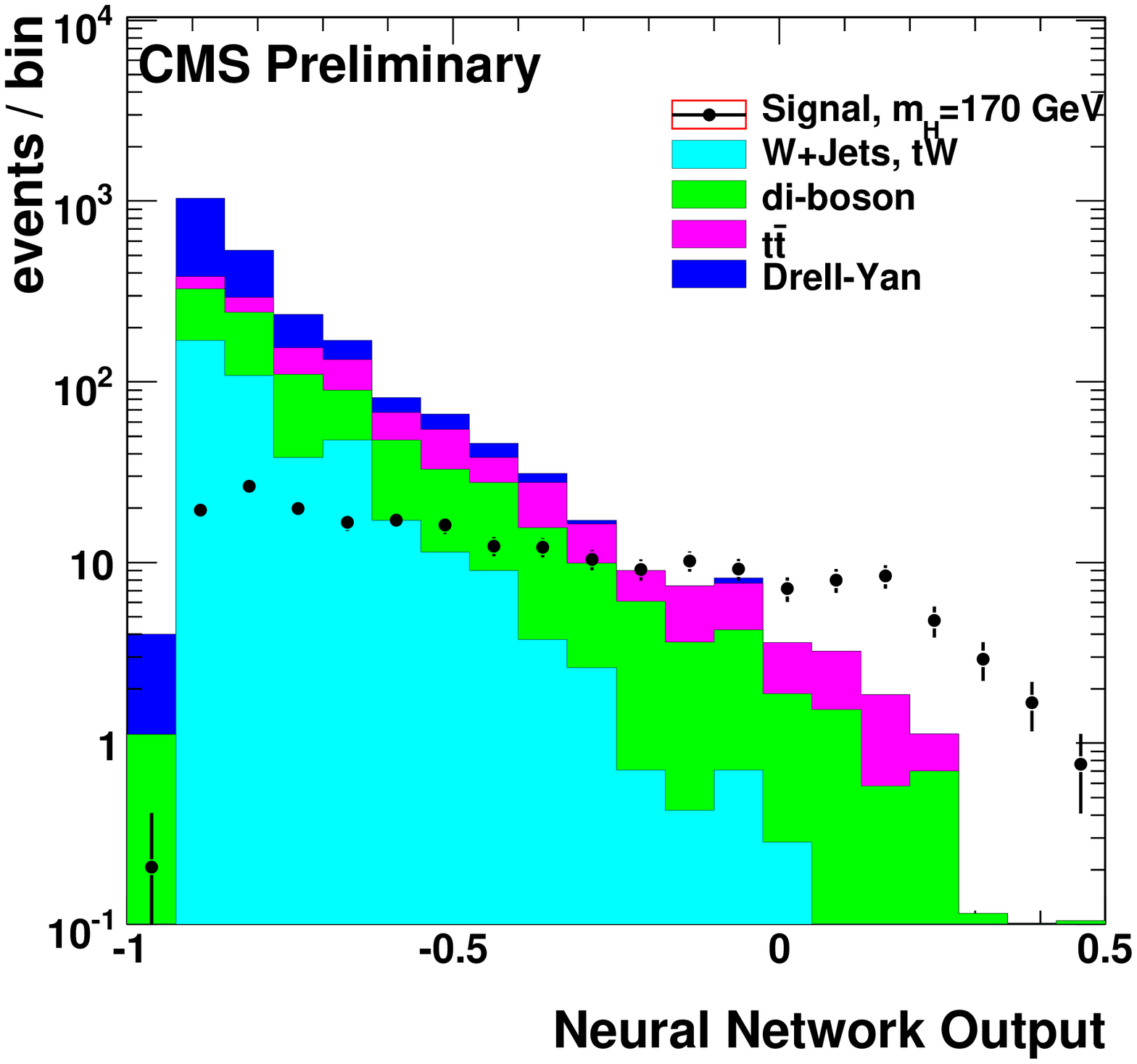,width=0.5\textwidth}
\psfig{figure=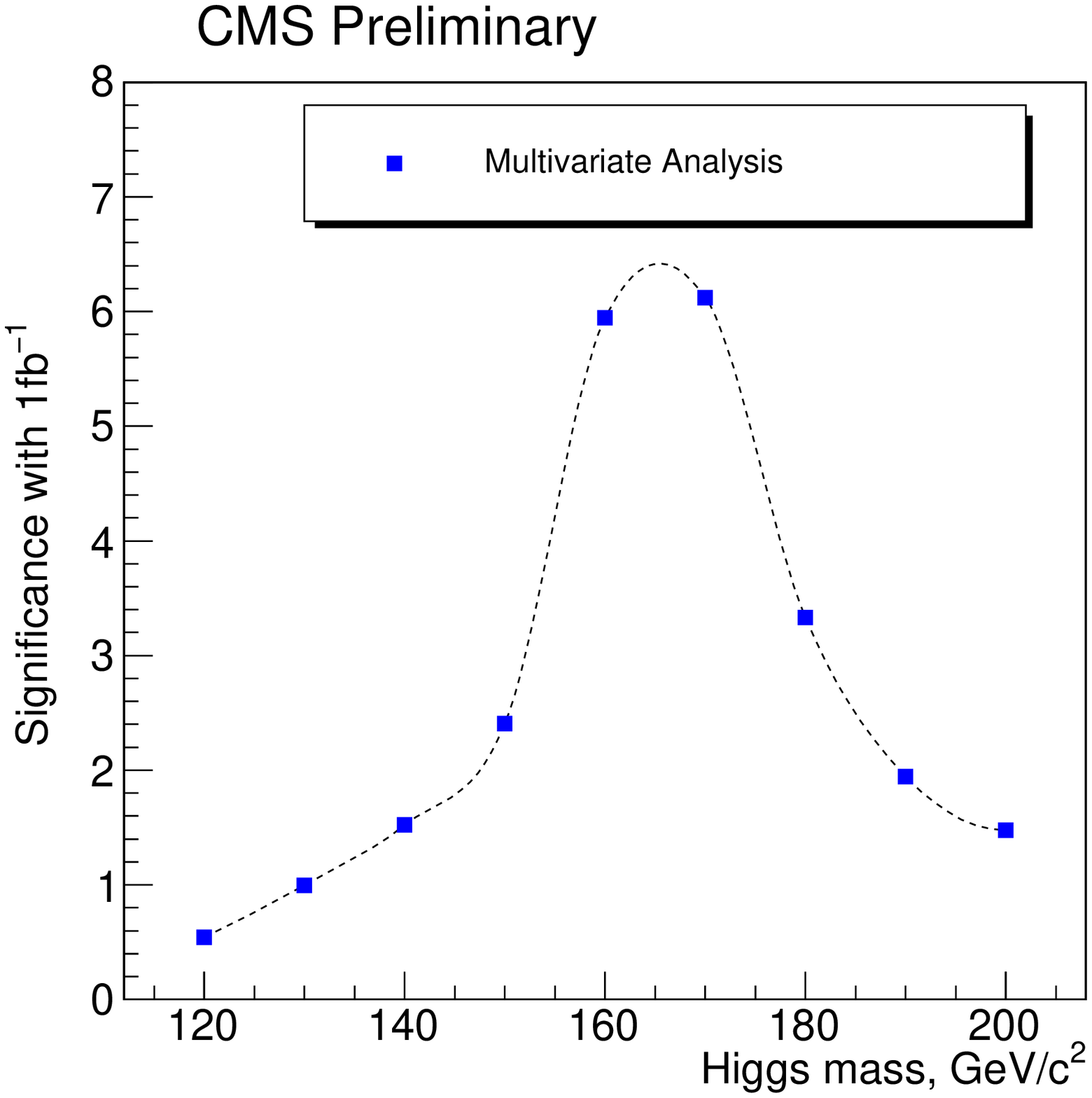,width=0.5\textwidth}
\caption{The distribution of the output result of the neural net (left) for signal and background in the $\mathrm{H \rightarrow WW \rightarrow 2l 2\nu}$ search, with $\mathrm{1 \, fb^{-1}}$ of integrated luminosity; significance of the signal observation (right) in the $\mathrm{H \rightarrow WW \rightarrow 2l 2\nu}$ with an integrated luminosity of $1 \,\mathrm{ fb^{-1}}$. }
\label{hww}
\end{figure}

Both a cut-based and neural net-based approaches were used to gain discrimination between signal and
background. The distribution of the output result of the neural net for signal and background is
reported in Fig.~\ref{hww} (left), for $\mathrm{1 \, fb^{-1}}$ of integrated luminosity. Strategies to control the efficiency of leptons and jet reconstruction, the rate of jets faking leptons, the measurement of
the missing energy and the estimation of $\mathrm{t \bar{t}}$ and WW
background rates from data were also developed.

The significance for the signal observation in CMS with $\mathrm{1 \, fb^{-1}}$ of integrated luminosity as a
function of the Higgs mass hypothesis is reported in Fig.~\ref{hww} (right);
that is converted in an equivalent number of one-sided tail $\mathrm{\sigma}$ of the Gaussian distribution
and it is larger than 3 for Higgs masses between 155 and 185 $\mathrm{GeV/c^2}$.

In the case of $\mathrm{H \rightarrow ZZ}$ decay channel the topology of four
leptons in the final state (electron and/or muons) was studied with
an integrated luminosity of 1 and 30 $\mathrm{fb^{-1}}$ for CMS~\cite{PAS_CMS} and ATLAS~\cite{ATLAS1} respectively; the irreducible background comes from the ZZ events with four leptons in the final
state while $Z\mathrm{ b \bar{b}}$ and $\mathrm{t \bar{t}}$ events
could be reduced.

A preselection strategy aimed to get rid of QCD
related background with jets faking leptons was developed in the CMS
collaboration; that is based on electron identification techniques,
loose isolation on leptons and a minimal cuts on di-lepton and
four-lepton invariant mass. $Z\mathrm{ b \bar{b}}$ and $\mathrm{t
\bar{t}}$ events were substantially reduced with a tight isolation
on leptons and cuts on their impact parameters at the closest
approach point. Another powerful observable is the mass of the
reconstructed off-mass shell Z. With the purpose of providing a
robust baseline strategy for the observation of the Higgs, the complete
selection is cut-based and $m_{H}$-independent. Strategies to control efficiencies of lepton reconstruction and estimate the rate of ZZ and $\mathrm{Z b \bar{b}}$ events from data were also developed.

\begin{figure}[t]
\psfig{figure=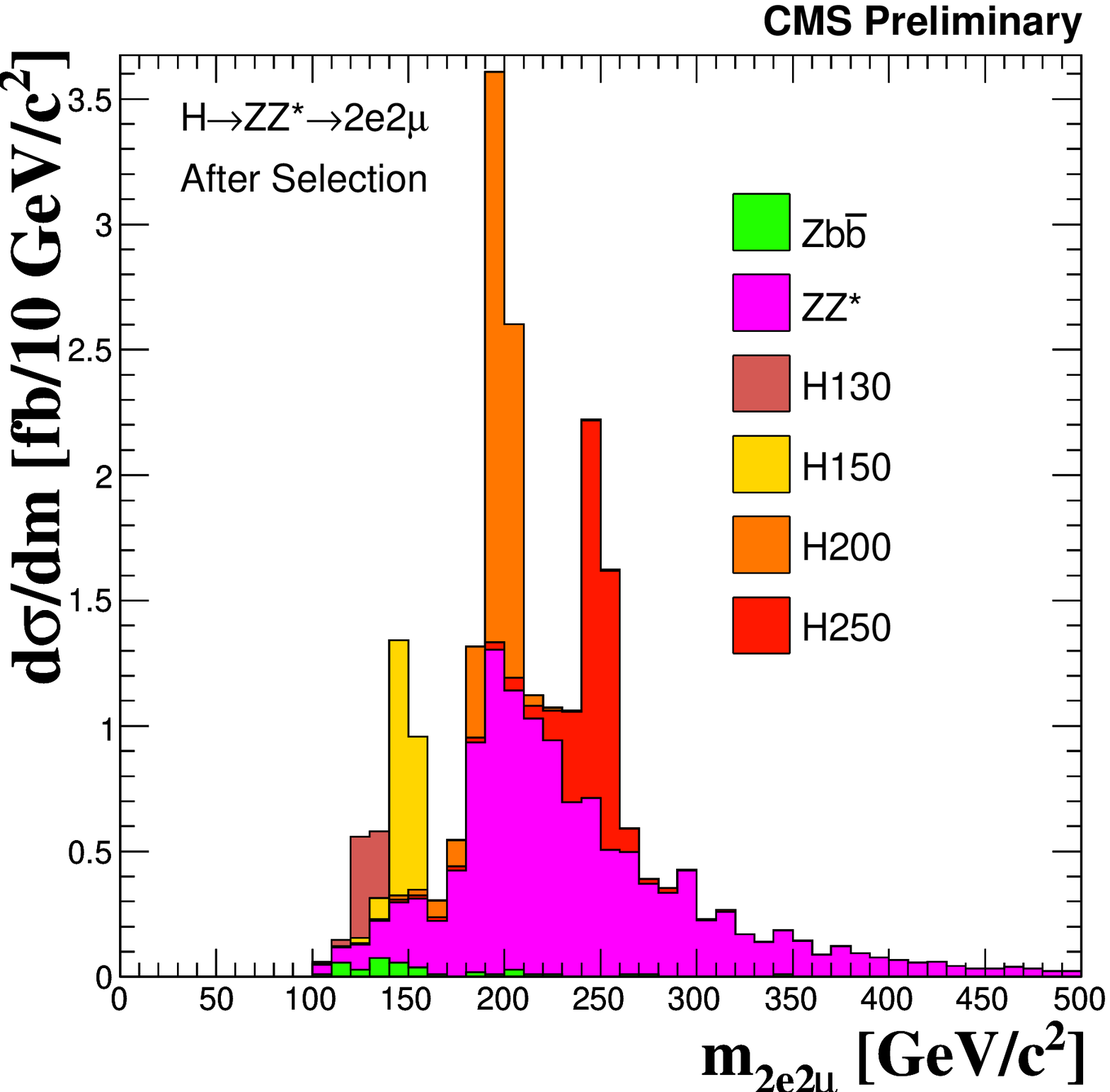,width=0.5\textwidth}
\psfig{figure=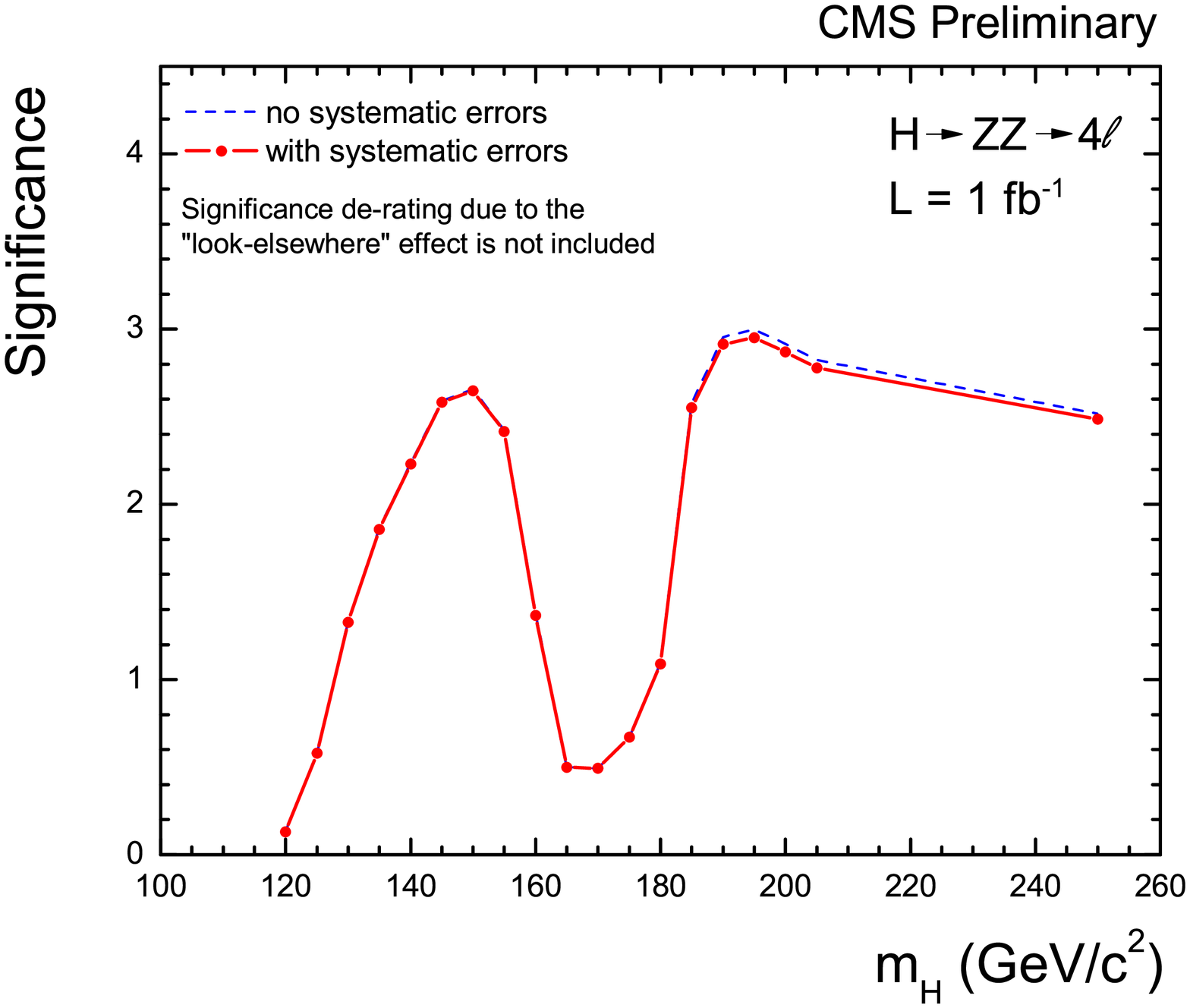, width=0.5\textwidth}
\caption{$\mathrm{2e2\mu}$ invariant mass (left) after the full selection, corresponding to an integrated luminosity of $1 \, \mathrm{ fb^{-1}}$; significance for the signal observation (right) in the $\mathrm{H \rightarrow ZZ \rightarrow 4l}$ channel with an integrated luminosity of $1 \, \mathrm{ fb^{-1}}$. }
\label{2e2mu}
\end{figure}

The four-lepton invariant mass spectrum obtained in the
case $2e 2\mu$ final state at the end of the selection is reported
in Fig.~\ref{2e2mu} (left). The significance for the signal observation with
an integrated luminosity of 1 $\mathrm{fb^{-1}}$ is reported
in Fig.~\ref{2e2mu} (right), as obtained by the CMS collaboration. The significance of
such an observation needs to be further de-rated by about 1s unit to
take into account the probability of a random fluctuation anywhere in the mass spectrum (the
so-called look-elsewhere effect); when taking into account that effect, it is unlikely that an integrated luminosity of
1 $\mathrm{fb^{-1}}$ will yield an observation of a mass peak with an overall significance above $\mathrm{2\sigma}$.

Even if the branching ratio of the decay in two photons $\mathrm{H
\rightarrow \gamma \gamma} $ is less than \% at low Higgs mass the
clear signature of the final state makes that topology very
promising. Background events come from the production of
two isolated photons, which are usually referred to as irreducible,
while reducible background sources are events with at least one fake photon.
Fake photons are mostly due to the presence of a leading $\mathrm{\pi^{0}}$
resulting from the fragmentation of a quark or a gluon.

\begin{figure}[t]
\psfig{figure=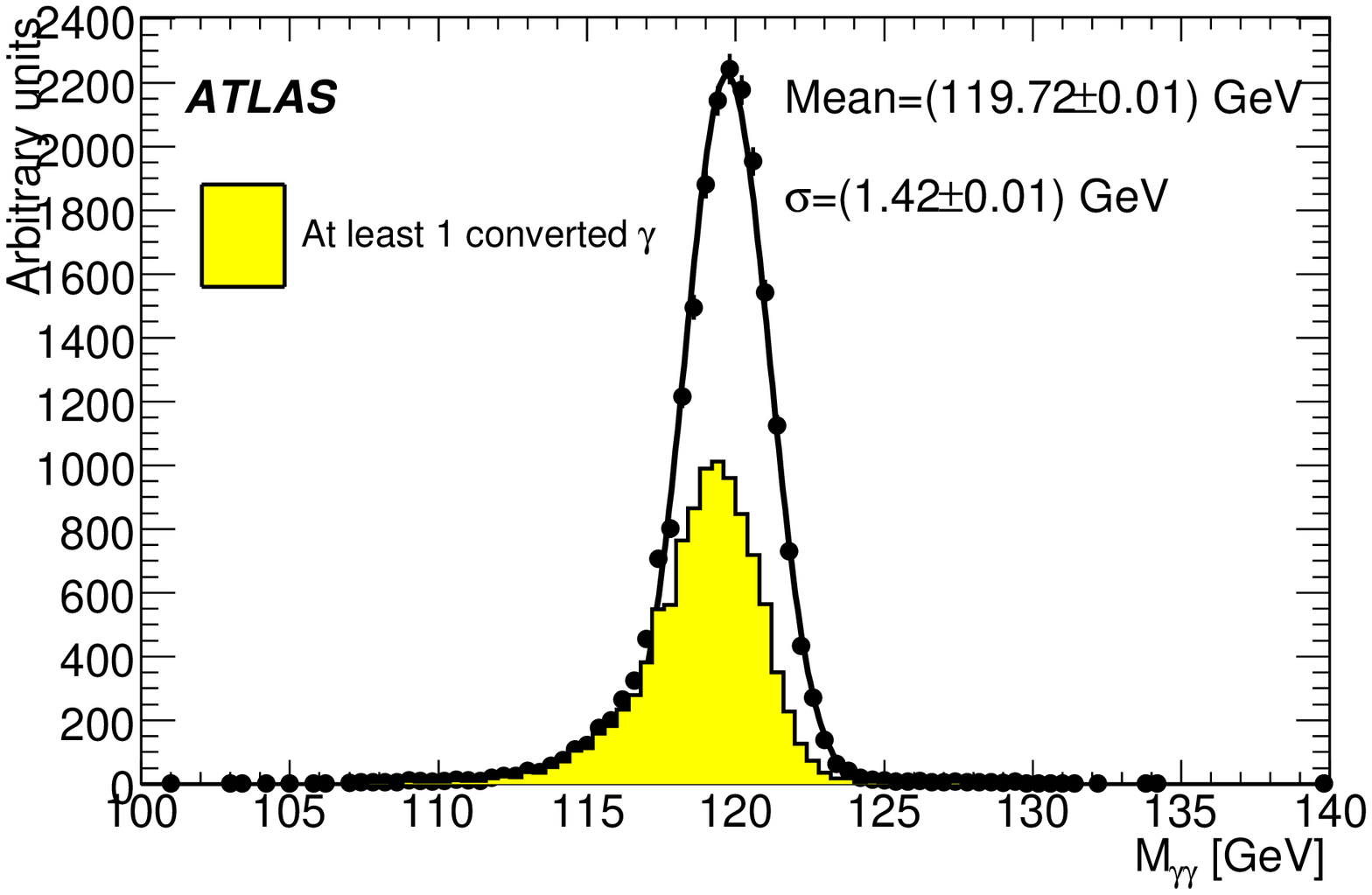,width=0.5\textwidth}
\psfig{figure=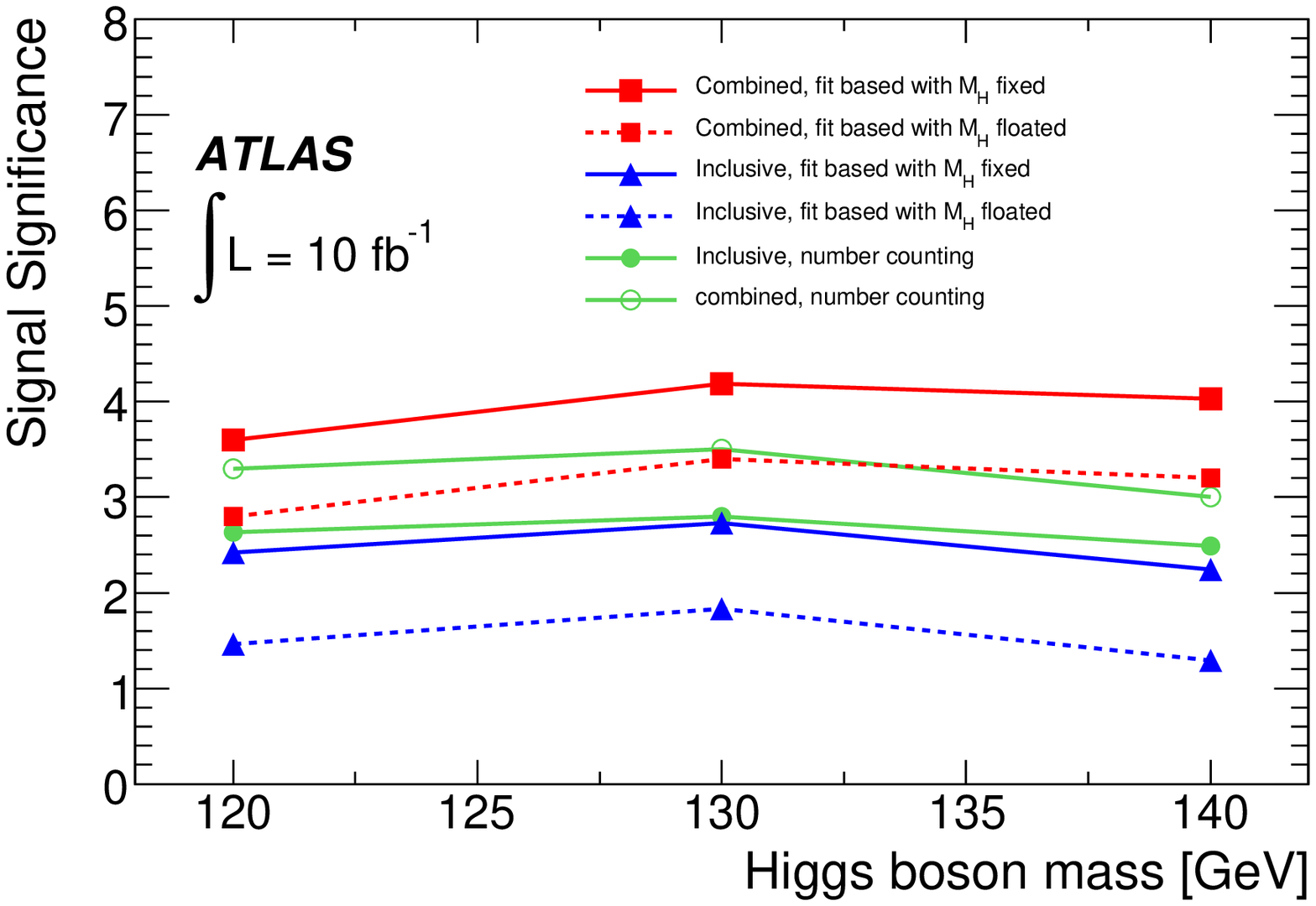,width=0.5\textwidth}
\caption{Invariant mass distributions (left) for photons pairs from Higgs boson decays with Higgs mass of 120 $\mathrm{GeV/c^2}$ after trigger and identification cuts; signal significance (right) in $\mathrm{H \rightarrow \gamma \gamma}$ channel as a function of the Higgs mass for 10 $\mathrm{fb^{-1}}$ of integrated luminosity . The solid circles correspond to the sensitivity of the inclusive analysis by using event counting. The open circles display the event counting significance when the Higgs boson plus jet analyses are included. The squares markers correspond to the sensitivity obtained using a combined analysis.}
\label{comb}
\end{figure}

The performance of the electromagnetic calorimeter and of the photon reconstruction, identification (to reject background from jets faking photons) and calibration are fundamental to disentangle the signal from the background. Considering Higgs boson decays with photons within the acceptance, about 57\% of the selected events have at least one true conversion with a radius smaller than 80 cm in the ATLAS detector.
Conversions are reconstructed by a vertexing algorithm using the reconstructed particle tracks.
Among the reconstructed photons passing the identification cuts, the two with highest transverse momentum are assumed to come from the Higgs boson decay so the vertex position of that is reconstructed. The invariant mass distributions for photons pairs from 120 $\mathrm{GeV/c^2}$ mass Higgs boson decays after trigger and identification cuts is reported in Fig.~\ref{comb} (left).

In the ATLAS collaboration, in addition to the inclusive $\mathrm{H \rightarrow \gamma \gamma}$ search, many topologies with one or two jets, with missing transverse energy
and isolated leptons or with only missing transverse energy, were also
studied \cite{ATLAS1}. The
significance in the $\mathrm{H \rightarrow \gamma \gamma}$ as a function
of the Higgs mass is reported in Fig.~\ref{comb} (right); a significance based on event counting
of 2.6 with 10 $\mathrm{fb^{-1}}$ for $\mathrm{m_{H} = 120 \, GeV}$ is obtained in the case of inclusive analysis.

\begin{figure}[t]
\psfig{figure=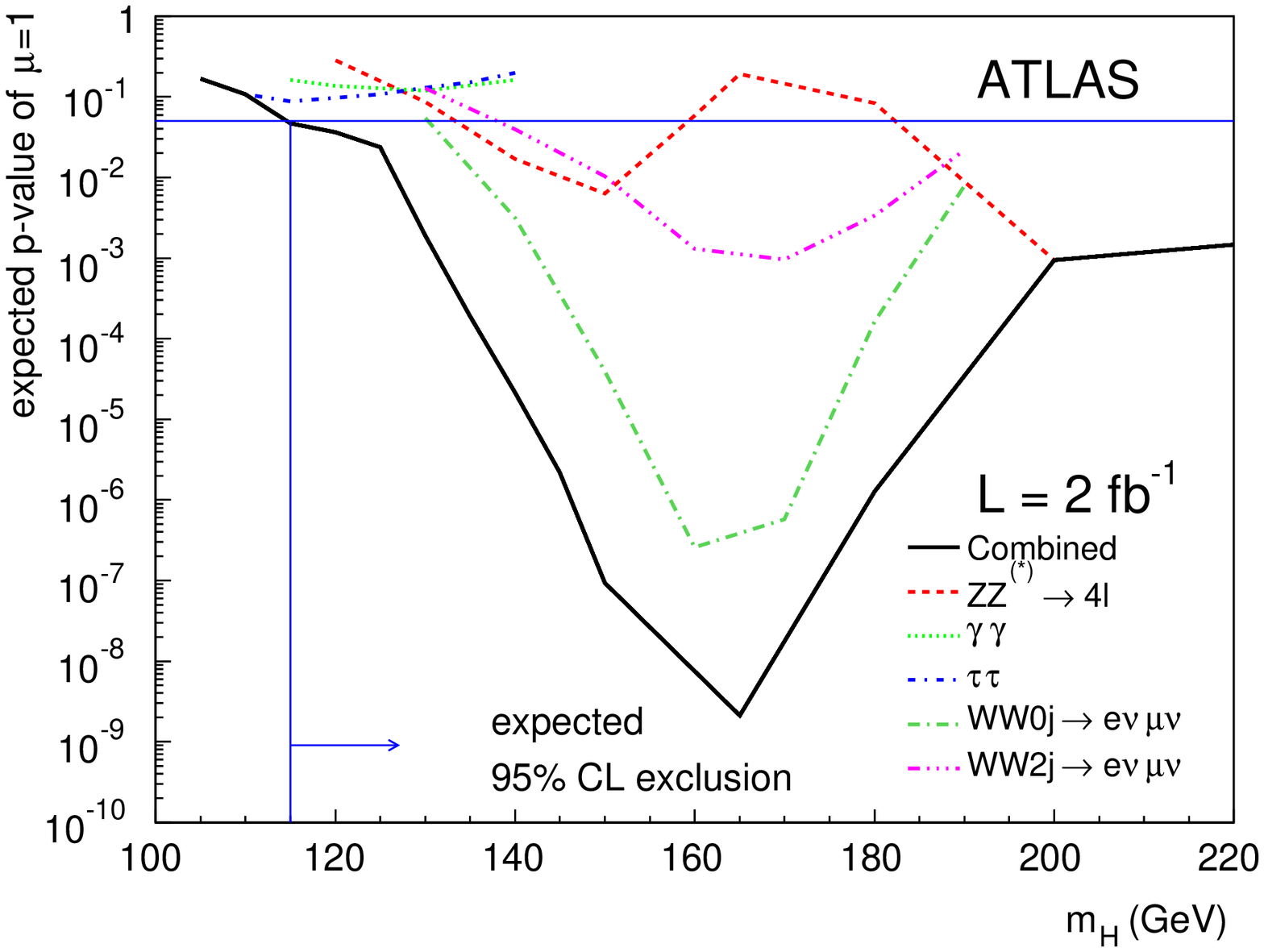,width=0.5\textwidth}
\psfig{figure=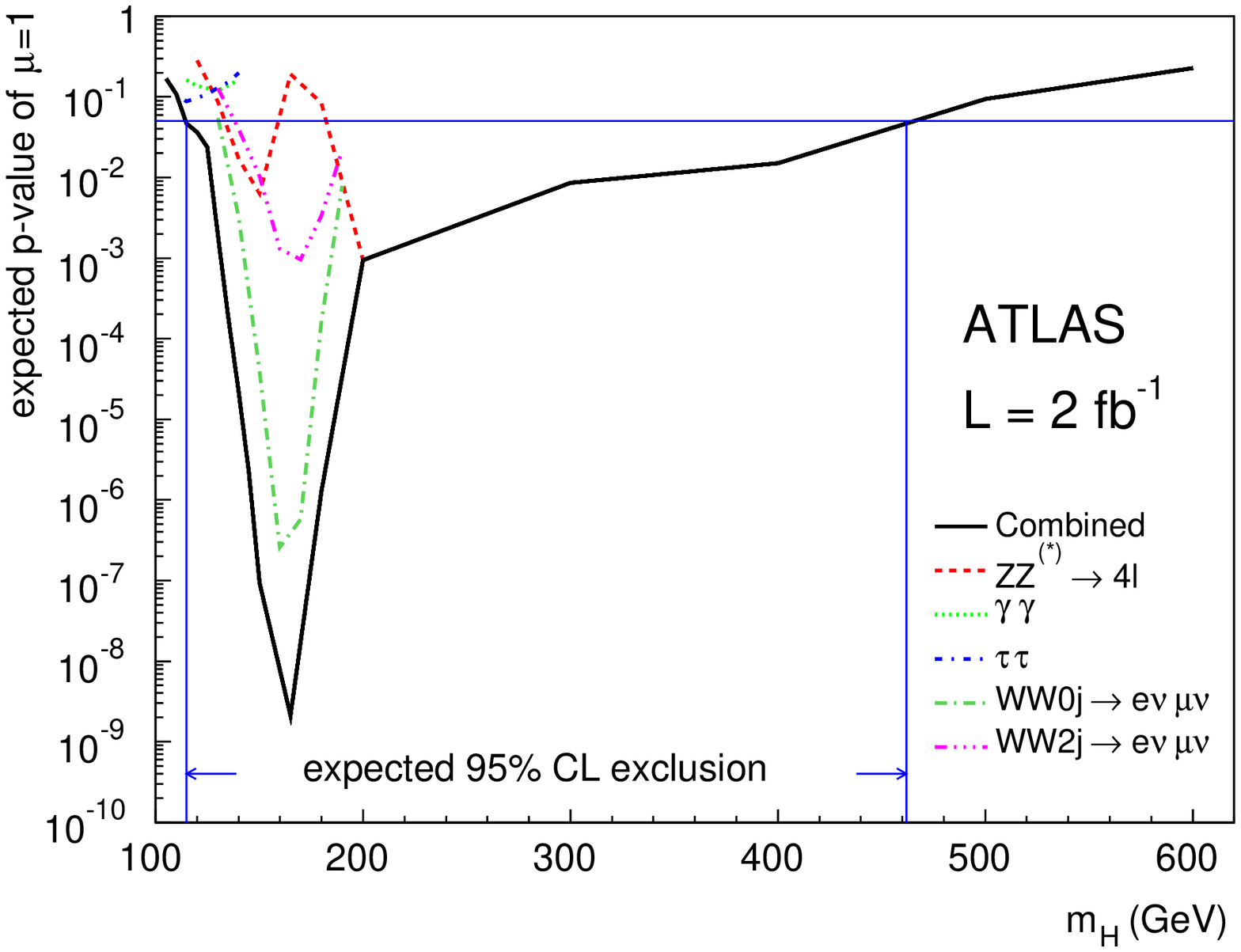,width=0.5\textwidth}
\caption{The median p-value obtained for excluding a SM Higgs Boson for the various channels
as well as the combination for (left) the lower mass range (right) for masses up to 600 GeV with integrated luminosity of 2 $\mathrm{fb^{-1}}$.}
\label{atlascomb}
\end{figure}

Statistical procedures for combination of results were used in the ATLAS collaboration to derive the potential of discovery and exclusion from independent searches: $\mathrm{H \rightarrow \tau^{+} \tau^{-}}$, $\mathrm{H \rightarrow WW \rightarrow e\nu \mu \nu} $, $\mathrm{H \rightarrow \gamma \gamma}$ and $\mathrm{H \rightarrow ZZ \rightarrow 4l }$, as detailed in Ref.~\cite{ATLAS1}. The level of compatibility between data that give an observed value of a given estimator (typically a likelihood ratio) and a given hypothesis (background only or signal+bagkround) is quantified by giving the p-value that is the probability, under the assumption of a given hypothesis, of seeing data with equal or greater incompatibility, relative to the data actually obtained. Any p-value below 0.05 indicates an exclusion; the median p-value obtained for excluding a SM Higgs Boson for the various channels as well as the combination with integrated luminosity of 2 $\mathrm{fb^{-1}}$ is reported in Fig.~\ref{atlascomb}; ATLAS has the median sensitivity to exclude a SM Higgs boson with a mass in a 115-460 GeV range at 95 \% C.L..

\subsection{Searches for supersymmetric particles}\label{subsec:susy}

Hints of supersymmetry \cite{MSSM} are looked for at LHC via the
production of squarks and gluinos, the supersymmetric partners of quark
and gluons of the SM. The final state topologies of the
supersymmetric events at LHC consist of multiple jets, often very
energetic, with possibly some leptons and missing energy in the
final state or simply with many leptons and missing energy.

Most of the studies performed in CMS and ATLAS were done in the context of
the Minimal Supersymmetric Standard Model (MSSM) with R-parity conservation
and in the scenario of heavy squarks and gluinos. in order to reduce
the number of free parameters of MSSM the hypotheses of minimal
Supergravity (mSUGRA~\cite{msugra}) are used, in particular by
assuming a common sfermion mass at GUT scale ($\mathrm{m_{0}}$) and
a common gaugino mass ($\mathrm{m_{1/2}}$).

Prospective analyses were developed to search for final states
including jets, leptons and missing energy both in CMS~\cite{susydijet} and in the ATLAS
collaboration~\cite{susyatlas}. Typically some benchmark points of the parameter space of MSSM with mSUGRA
hypotheses are used as starting points and scans of parameters around them is performed to
derive conservative limits.

Concerning ATLAS analyses, one possible inclusive signature is consist of four jets and missing energy. The
main backgrounds are $\mathrm{t \bar{t}}$ and W/Z+jets events. Simple selection
cuts are applied on the total transverse momentum of the jets, on the missing
transverse energy, on the angle between the jet and the missing energy directions
and on the effective mass of transverse momentum of the jets and leptons and
 missing transverse energy. Final state topologies with less than four jets and with one or
 more leptons were also studied.

\begin{figure}[thp]
\psfig{figure=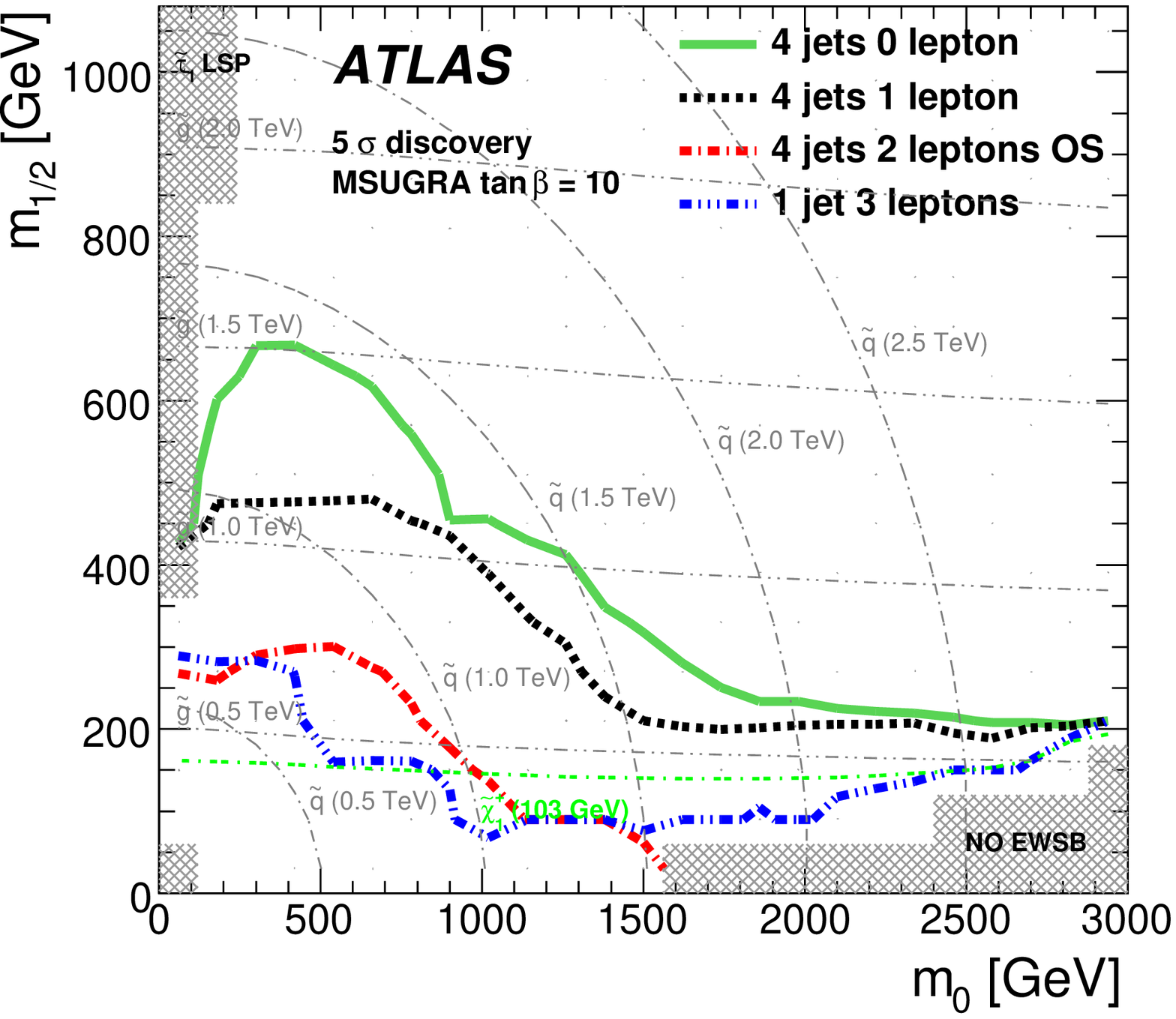,width=0.5\textwidth}
\psfig{figure=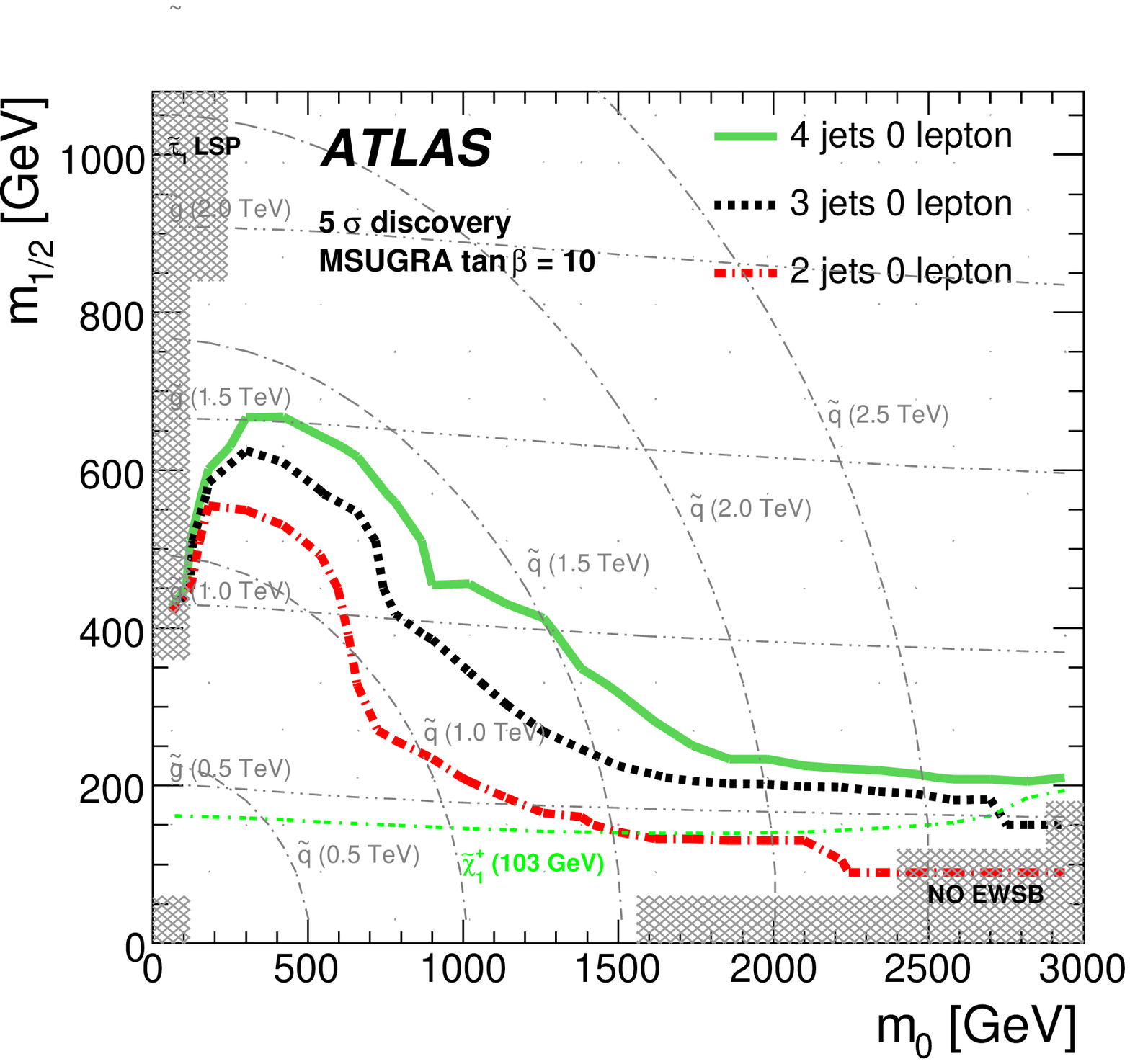,width=0.5\textwidth}
\caption{The $\mathrm{5 \sigma}$ discovery reach  in the plane
($\mathrm{m_{0},m_{1/2}}$) in the case of four jets with one or more
leptons in the final state and missing energy (left) and in the case
of two, three and four jets with zero lepton (right).}
\label{susy}
\end{figure}

In the Fig.~\ref{susy} (left) is reported the $\mathrm{5 \sigma}$ discovery reach in the plane
($\mathrm{m_{0}},m_{1/2}$) in the case of four jets with one or more
leptons in the final state and missing energy; zero-lepton mode can probe close to 1.5 TeV
for the minimum between the squark and the gluino mass, with 1
$\mathrm{fb^{-1}}$ of integrated luminosity; the four-jets topology seems to give the best results in
zero-lepton mode, as derived by Fig.~\ref{susy} (right). Therefore ATLAS could discover signals
with gluino and squark masses less than O(1 TeV) after having accumulated an
integrated luminosity of about $\mathrm{1 \,fb^{-1}}$.

\subsection{Searches for exotic particles}\label{subsec:extrad}

Exotic massive gauge bosons are expected in several theoretical models
beyond the SM. In the sequential Standard Model~\cite{SSM} (SSM) a Z-like boson,
called Z', with the same couplings of the
Z to fermions and gauge bosons and with O(TeV) mass is predicted.
Other exotic scenarios based on extra dimension~\cite{RS} predict
the existence of a graviton with O(TeV) mass decaying in
$\mathrm{e^+ e^-}$.

Searches for high mass gauge bosons decaying in $\mathrm{e^+ e^-}$
pair were performed both in CMS~\cite{gravitonee} and ATLAS~\cite{ATLAS1}. The cross section times
the branching ratio is between few fb to few hundred fb depending on
the mass of the resonance and the theoretical model. Main
backgrounds for those searches were di-electron events produced via
Drell Yan mechanism, $\mathrm{t \bar{t}}$ events with two electrons
in the final state, QCD with jets faking electrons, W+jets,
$\mathrm{\gamma}$+jets, $\mathrm{\gamma \gamma}$.

Concerning the CMS analysis, an important aspect of the analysis was the usage of
high threshold trigger patterns to tag those events ($\mathrm{E_{T}>
80 \, GeV}$ and loose isolation on leptons with $\mathrm{E_{T}>200 \,
GeV}$ in electromagnetic calorimeter). Saturation occurs in the electromagnetic calorimeter
electronics for very high energy deposits in a single ECAL crystal
($\mathrm{> \, 1.7 \, TeV}$ for the barrel and $ \mathrm{> \, 3.0 \, TeV}$ for the endcaps); the energy in the saturated
crystal can be reconstructed, with a resolution of about 7\%, using the energy deposit
distribution in the surrounding crystals, as detailed in Ref.~\cite{saturation}.

\begin{figure}[t]
\psfig{figure=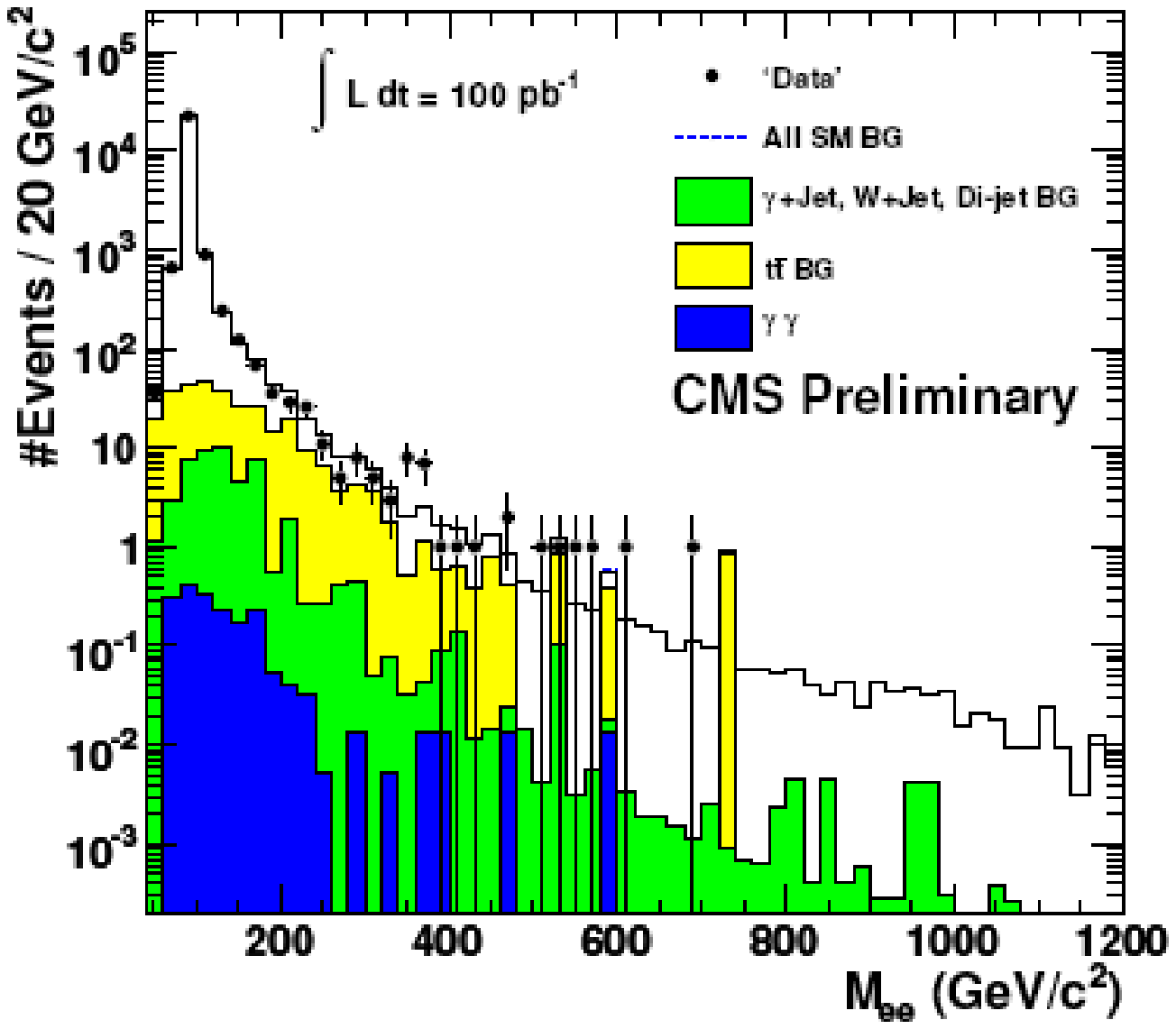,width=0.5\textwidth}
\psfig{figure=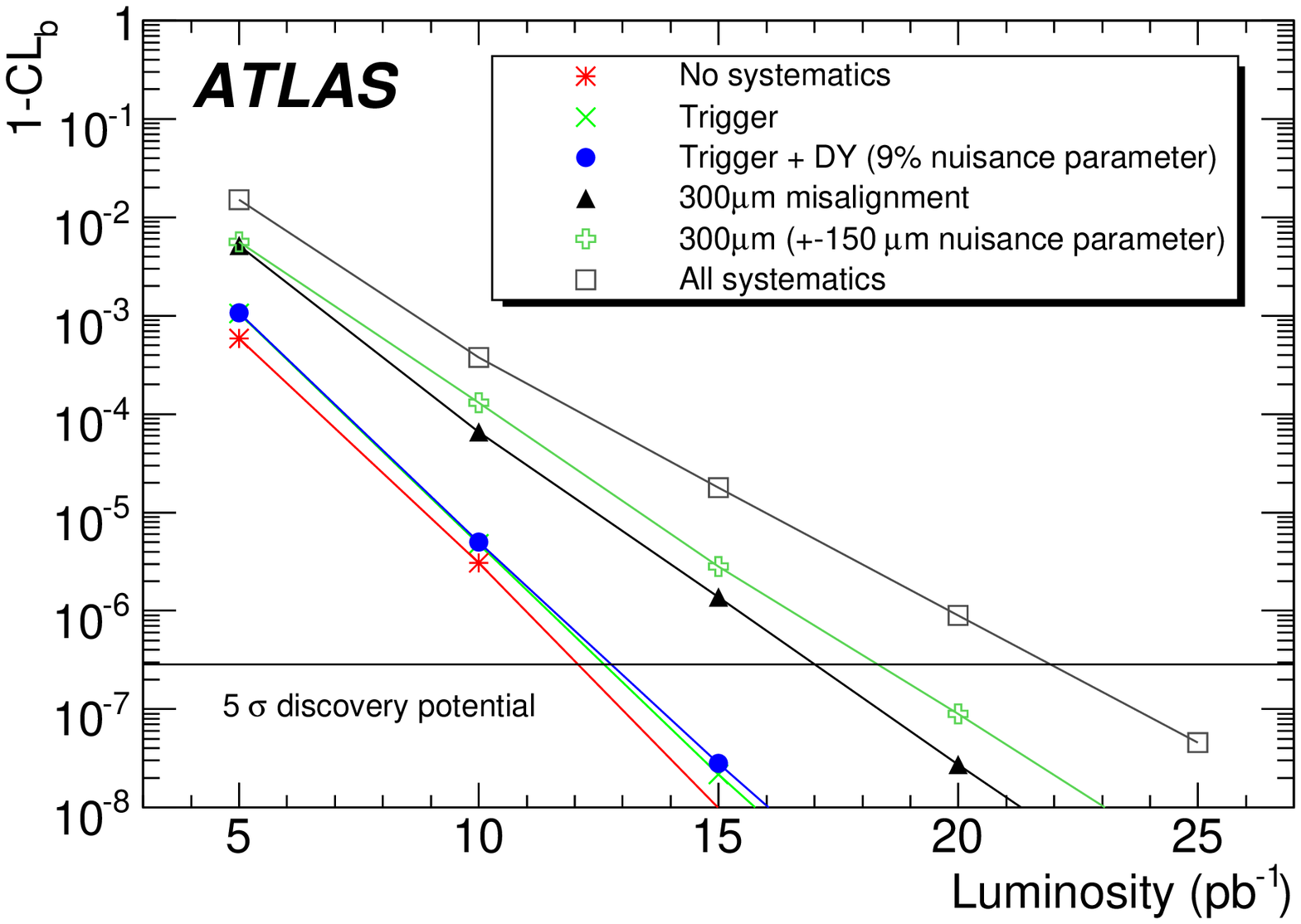,width=0.5\textwidth}
\caption{Di-electron invariant mass spectrum (left) for a 100 $\mathrm{pb^{-1}}$ integrated luminosity
with 1 $\mathrm{TeV/c^2}$ Z' signal, compared to SM background estimates for the Drell-Yan process, $\mathrm{t \bar{t}}$, QCD
di-jet, W+jet, $\mathrm{\gamma}$+jet and $\mathrm{\gamma \gamma}$; $\mathrm{1-CL_{b}}$ distribution (right) obtained as a function of the integrated luminosity for the Z' expected in the SSM at mass of 1 $\mathrm{TeV/c^2}$, if the muon spectrometer is aligned with a precision of 300 $\mathrm{\mu m}$. The effect of the systematic uncertainty on the trigger selection and on the knowledge of the SM Drell-Yan cross-section is also displayed.}
\label{zprime}
\end{figure}

The di-electron invariant mass spectrum for signal and background
at 100 $\mathrm{pb^{-1}}$ is reported in Fig.~\ref{zprime} (left); at high mass only
few background events survive the selection giving an optimal signal to
background rejection.

At the end of the analysis, after computing the integrated luminosity for
5$\mathrm{\sigma}$ discovery at $\mathrm{\sqrt s = 14\, TeV}$ as a function of the Z' mass,
it could be shown that few hundred $\mathrm{pb^{-1}}$ of integrated
luminosity are needed to discover the Z' with O(1 TeV) mass with 5$\mathrm{\sigma}$.

Search for di-muon resonances at O(1TeV) mass were addressed too by
CMS and ATLAS~\cite{ATLAS1}. Sources of background are di-muons from Drell Yan
events and W+jets, Z+jets.


At large transverse momentum ($ \mathrm{> \, 100 \, GeV}$), an important contribution to the muon momentum resolution
is related to the misalignment of the muon spectrometer. A detailed study was carried out in order
to determine the effect of possible larger uncertainties in the position of the muon chambers to the Z' search;
in addition to the ideal case of no misalignment, several different hypotheses of misalignment were simulated.
Muon chamber misalignment has an important effect causing a loss of Z' mass resolution that degrade
the determination of the charge of muon.

In the Fig.~\ref{zprime} (right) is reported the luminosity needed for a
5$\mathrm{\sigma}$ discovery of Z' as predicted by the SSM. That luminosity ranges from 20 to 40 $\mathrm{pb^{-1}}$,
which makes the di-muon channel competitive with the di-electron
channel. The inclusion of the effect of misalignment and all the
systematics makes the prediction less powerful and the result worst.

%
%
%

\end{document}